\documentclass[12pt]{amsart}
\usepackage{amssymb}
\usepackage{color}
\usepackage{amsmath,epic,curves}
\usepackage[english]{babel}
\usepackage{graphicx}
\newtheorem{claim}{}[section]
\newtheorem{theorem}[claim]{Theorem}

\begin{document}
\baselineskip 6.0 truemm
\parindent 1.5 true pc

\newcommand\lan{\langle}
\newcommand\ran{\rangle}
\newcommand\tr{{\text{\rm Tr}}\,}
\newcommand\ot{\otimes}
\newcommand\ol{\overline}
\newcommand\join{\vee}
\newcommand\meet{\wedge}
\renewcommand\ker{{\text{\rm Ker}}\,}
\newcommand\image{{\text{\rm Im}}\,}
\newcommand\id{{\text{\rm id}}}
\newcommand\tp{{\text{\rm tp}}}
\newcommand\pr{\prime}
\newcommand\e{\epsilon}
\newcommand\la{\lambda}
\newcommand\inte{{\text{\rm int}}\,}
\newcommand\ttt{{\text{\rm t}}}
\newcommand\spa{{\text{\rm span}}\,}
\newcommand\conv{{\text{\rm conv}}\,}
\newcommand\rank{\ {\text{\rm rank of}}\ }
\newcommand\re{{\text{\rm Re}}\,}
\newcommand\ppt{\mathbb T}
\newcommand\rk{{\text{\rm rank}}\,}
\newcommand\sss{\omega}

\title{Global geometric difference between separable and Positive partial transpose states}

\author{Kil-Chan Ha}
\address{Faculty of Mathematics and Statistics, Sejong University, Seoul 143-747, Korea}

\author{Seung-Hyeok Kye}
\address{Department of Mathematics and Institute of Mathematics\\Seoul National University\\Seoul 151-742, Korea}

\thanks{KCH is partially supported by NRFK 2013-020897. SHK is partially supported by NRFK 2009-0083521}



\begin{abstract}
In the convex set of all $3\ot 3$ states with positive partial
transposes, we show that one can take two extreme points whose
convex combinations belong to the interior of the convex set. Their
convex combinations may be even in the interior of the convex set of
all separable states. In general, we need at least $mn$ extreme
points to get an interior point by their convex combination, for the case of the convex set of
all $m\ot n$ separable states. This shows a sharp distinction
between PPT states and separable states. We also consider the same
questions for positive maps and decomposable maps.
\end{abstract}

\maketitle

\section{Introduction}

Distinguishing entanglement from separability is one of the most
important question in the theory of quantum entanglement, and the
positive partial transpose (PPT) criterion \cite{choi-ppt,peres}
gives a simple but strong necessary condition for separability. The
PPT condition is actually equivalent to the separability if the rank
of a given PPT state is sufficiently low by \cite{hlvc}. In the case
that the rank is not so high, it turns out that the local geometry is
quite useful to distinguish and construct entanglement among PPT
states. Basic idea is to consider the smallest faces determined by a
given separable state in the convex sets of all separable states and
all PPT states respectively, and compare those. See
\cite{ha+kye_unique_decom,ha+kye_2x4} for recent progresses in this
direction.

In this paper, we turn our attention to the global geometries for separable and PPT states,
and look for the differences.
We denote by $\mathbb S_{m,n}$ the convex set of all $m\otimes n$ separable states,
and by $\mathbb T_{m,n}$ the convex set of all $m\ot n$ PPT states.
For the convex set $\mathbb S_{m,n}$, it is easy to see that a convex combination
of two extreme points is always on the boundary of the convex set. Actually, the line
segment between two extreme points of $\mathbb S_{2,2}$ is already a nontrivial face of the convex set, in most cases.
See \cite{kye_trigono} for more details for the convex geometry of $\mathbb S_{2,2}$.
More generally, the convex hull of $\max\{m,n\}$ extreme points of $\mathbb S_{m,n}$
is a face of the convex set, in most cases by \cite{alfsen,kirk}.
Therefore, it is natural to ask how these properties are retained for the convex set $\mathbb T_{m,n}$.

For a convex compact set $C$ in a finite dimensional real vector space,
we introduce the number $\nu(C)$ the smallest natural number $k$ such that the convex combination
of $k$ extreme points of $C$ may be an interior point of $C$. We recall that the interior of a convex set
is defined by the interior with respect to the relative topology induced by the affine manifold generated by itself.
Sometimes, it is more convenient to consider the convex cone $\tilde C$ generated by the convex compact set $C$.
For example, $\tilde{\mathbb  S}_{m,n}$ ($\tilde{\mathbb T}_{m,n}$, respectively)
is the convex cone of all $m\otimes n$ unnormalized separable (PPT, respectively) states.
In this case, we may replace extreme points by extreme rays to get the same number $\nu(C)$.
Recall that a point $x$ of a convex compact set $C$ is an extreme point of $C$ if and only if $x$ generates an extreme ray of
the convex cone $\tilde C$, whenever the hyperplane generated by $C$ does not contain the origin.

It is easy to see that
$$
\nu(\mathbb S_{m,n})=mn,
$$
for every $m,n=2,3,\dots$. The main purpose of this note is to show that
$$
\nu(\mathbb T_{3,3})=2,
$$
to see the geometric difference between $\mathbb S_{3,3}$ and $\mathbb T_{3,3}$.
In other words, we can take just two extreme points of $\mathbb T_{3,3}$ whose convex combinations
belong to the interior of $\mathbb T_{3,3}$.

To do this, we consider the following $3\otimes 3$ states
$$
\varrho_{b,\theta}
=\left(
\begin{array}{ccccccccccc}
p_\theta     &\cdot   &\cdot  &\cdot  &-e^{i\theta}     &\cdot   &\cdot   &\cdot  &-e^{-i\theta}    \\
\cdot   &\frac 1b &\cdot    &-e^{-i\theta}    &\cdot   &\cdot &\cdot &\cdot     &\cdot   \\
\cdot  &\cdot    &b &\cdot &\cdot  &\cdot    &-e^{i\theta}    &\cdot &\cdot  \\
\cdot  &-e^{i\theta}    &\cdot &b &\cdot  &\cdot    &\cdot    &\cdot &\cdot  \\
-e^{-i\theta}     &\cdot   &\cdot  &\cdot  &p_\theta     &\cdot   &\cdot   &\cdot  &-e^{i\theta}     \\
\cdot   &\cdot &\cdot    &\cdot    &\cdot   &\frac 1b &\cdot &-e^{-i\theta}    &\cdot   \\
\cdot   &\cdot &-e^{-i\theta}    &\cdot    &\cdot   &\cdot &\frac 1b &\cdot    &\cdot   \\
\cdot  &\cdot    &\cdot &\cdot &\cdot  &-e^{i\theta}    &\cdot    &b &\cdot  \\
-e^{i\theta}     &\cdot   &\cdot  &\cdot  &-e^{-i\theta}     &\cdot   &\cdot   &\cdot  &p_\theta
\end{array}
\right)
$$
for a given positive number $b>0$ and a real number $\theta$, where $\cdot$ denotes zero and
$$
p_\theta=\max\{ e^{i(\theta-\frac 23 \pi)}+e^{-i(\theta-\frac23 \pi)},
e^{i\theta}+e^{-i\theta},
e^{i(\theta+\frac 23 \pi)}+e^{-i(\theta+\frac 23 \pi)}\}
$$
is the smallest positive number $a$ so that the following $3\times
3$ matrix
$$
\left(
\begin{matrix}
a & -e^{i\theta} & -e^{-i\theta}\\
-e^{-i\theta} & a & -e^{i\theta}\\
-e^{i\theta} & -e^{-i\theta} & a
\end{matrix}
\right)
$$
is positive, as it was discussed in Section 2 of \cite{ha+kye_Choi}.
We note that $1\le p_\theta\le 2$. Therefore, it is immediate to see that $\varrho_{b,\theta}$ is a PPT state.
These PPT states have been constructed in \cite{kye_osaka} for $-\pi/3<\theta<\pi/3$. The main point is to
extend this construction for the full range of $\theta$.

We check that they are extreme points of $\mathbb T_{3,3}$ in most cases, with a few exceptions.
Note that the case of $b=2$ and $\theta=\pi/6$ has been checked to be extreme in \cite{chen_dj_3x3}.
If we divide the parameter $e^{i\theta}$ into three arcs and take any two extreme points from different arcs then their
convex combinations lie in the interior of $\mathbb T_{3,3}$. We see that some of them turn out to be even
in the interior of $\mathbb S_{3,3}$.
It had been asked in \cite{chen_dj_extreme_ppt} whether the sum of two PPT entangled extreme states can be separable,
and the authors \cite{ha+kye_unique_decom} gave an affirmative answer. More precisely, it was shown that
sum of two extreme PPT entangled states with rank four may be separable. Our construction in this paper shows that
sum of two extreme PPT entangled states with rank five may be even diagonal matrices with positive diagonal entries.

Let $M_n$ be the $C^*$-algebra consisting of all $n\times n$ matrices over the complex field.
We also consider the same question for the convex cone $\mathbb P_{m,n}$ (respectively $\mathbb D_{m,n}$) of all
positive maps (respectively decomposable positive maps) from $M_m$ into $M_n$, to show that
$\nu(\mathbb D_{m,n})\ge m+n-2$.
In the case of $n=m=3$, we have $\nu(\mathbb D_{3,3})=4$ and $\nu(\mathbb P_{3,3})=2$.
By the Jamio\l kowski-Choi isomorphism \cite{choi75-10,jami}, the cones $\mathbb P_{m,n}$ and $\mathbb D_{m,n}$
are considered as subsets of $M_m\ot M_n$, and we have the relation
$$
\tilde{\mathbb S}_{m,n}\subset \tilde{\mathbb T}_{m,n}\subset \mathbb D_{m,n}\subset\mathbb P_{m,n}.
$$
We also recall \cite{eom-kye} that $\tilde{\mathbb S}_{m,n}$ and $\mathbb P_{m,n}$ (respectively $\tilde{\mathbb T}_{m,n}$ and $\mathbb D_{m,n}$)
are dual to each others  with respect to the bilinear pairing
\begin{equation}\label{pairing}
\langle \rho,\phi\rangle=\text{\rm Tr}(\rho C_{\phi}^{\rm t}),\qquad
\rho\in \tilde{\mathbb S}_{m,n},\  \phi \in \mathbb P_{m,n},
\end{equation}
where $C_{\phi}$ is the Choi matrix of $\phi$ defined by
$\sum |i\rangle \langle j|\otimes \phi(|i\rangle \langle j|)$, and  interior points of these convex sets can be characterized by the above duality:
\begin{itemize}
\item $\rho$ is an interior point of $\mathbb S_{m,n}$ if and only if $\langle \rho,\phi\rangle >0$ for all nonzero $\phi\in \mathbb
P_{m,n}$.
\item $\phi$ is an interior point of $\mathbb P_{m,n}$ if and only if $\langle \rho,\phi\rangle>0$ for all $\rho\in \mathbb
S_{m,n}$.
\end{itemize}
In this characterization of interior points of convex sets, we note that
it suffices to check the positivity of the pairing only for extreme points (rays) of the dual convex set (convex cone).
See Proposition 5.1 and 5.4 of \cite{kye_ritsu}.

In the next section, we examine the properties of the states $\varrho_{b,\theta}$ and how to choose two of them
to get an interior point by their convex combination. In Section 3, we show that they are extreme points in
$\mathbb T_{3,3}$, in most cases. We consider the convex cones $\mathbb P_{3,3}$ and $\mathbb D_{m,n}$ in Section 4,
and close this note with discussions in Section 5.

\section{Separable states and PPT states}

The facial structures for the convex set $\mathbb T_{m,n}$ are well understood \cite{ha_kye_04}.
Every face of $\mathbb T_{m,n}$ is of the form
$$
\tau(D,E)=\{\varrho\in\mathbb T_{m,n}: {\mathcal R}\varrho\subset D,\ {\mathcal R}\varrho^\Gamma \subset E\},
$$
for subspaces $D$ and $E$ of $\mathbb C^m\ot\mathbb C^n$, and its interior is given by
$$
\inte\tau(D,E)=\{\varrho\in\mathbb T_{m,n}: {\mathcal R}\varrho= D,\ {\mathcal R}\varrho^\Gamma = E\},
$$
where ${\mathcal R}\varrho$ denotes the range space of $\varrho$, and $\varrho^\Gamma$ is the partial
transpose of $\varrho$. Especially, a PPT state $\varrho$ is an interior point
of $\mathbb T_{m,n}$ if and only if the ranges of $\varrho$ and $\varrho^\Gamma$ are full spaces.

Extreme points of the convex set $\mathbb S_{m,n}$ are nothing but product states by the definition of separability.
If we take $k$ product states with $k<mn$ and form a separable state $\varrho\in\mathbb S_{m,n}$ with their
convex combination then
the range space of $\varrho$ is never the full space, and so $\varrho$ is on the boundary of $\mathbb T_{m,n}$.
By the relation $\mathbb S_{m,n}\subset\mathbb T_{m,n}$, we conclude that $\varrho$ is also on the boundary
of $\mathbb S_{m,n}$. Therefore, we have $\nu(\mathbb S_{m,n})\ge mn$. Since the identity matrix is in the interior
of $\mathbb S_{m,n}$, we conclude that $\nu(\mathbb S_{m,n})= mn$.
In fact, it is easy to see that every diagonal matrix with nonzero positive diagonal entries is an interior point
of the convex set $\mathbb S_{m,n}$, by the duality between separable states and positive maps.

Now, we proceed to examine the properties of the states $\varrho_{b,\theta}$. We also consider the PPT states
defined by
$$
\sigma_{b,\theta}=
\left(
\begin{array}{ccccccccccc}
p_\theta     &\cdot   &\cdot  &\cdot  &-e^{i\theta}     &\cdot   &\cdot   &\cdot  &-e^{-i\theta}    \\
\cdot   &\frac 1b &\cdot    &\cdot    &\cdot   &\cdot &\cdot &\cdot     &\cdot   \\
\cdot  &\cdot    &b &\cdot &\cdot  &\cdot    &\cdot    &\cdot &\cdot  \\
\cdot  &\cdot    &\cdot &b &\cdot  &\cdot    &\cdot    &\cdot &\cdot  \\
-e^{-i\theta}     &\cdot   &\cdot  &\cdot  &p_\theta     &\cdot   &\cdot   &\cdot  &-e^{i\theta}     \\
\cdot   &\cdot &\cdot    &\cdot    &\cdot   &\frac 1b &\cdot &\cdot    &\cdot   \\
\cdot   &\cdot &\cdot    &\cdot    &\cdot   &\cdot &\frac 1b &\cdot    &\cdot   \\
\cdot  &\cdot    &\cdot &\cdot &\cdot  &\cdot    &\cdot    &b &\cdot  \\
-e^{i\theta}     &\cdot   &\cdot  &\cdot  &-e^{-i\theta}     &\cdot   &\cdot   &\cdot  &p_\theta
\end{array}
\right).
$$
If $0<|\theta|<\frac\pi 3$
then $\sigma_{b,\theta}$ is nothing but PPT entangled edge states of type $(8,6)$ constructed in \cite{kye_osaka}.
We recall that a PPT state $\varrho$ is said to be of type $(p,q)$ if the ranks of $\varrho$ and $\varrho^\Gamma$ are $p$ and $q$,
respectively. We note that the state $\varrho_{b,\theta}$ defined in Introduction is given by
$$
\varrho_{b,\theta}
=\sigma_{b,\theta}+\sigma_{b,\theta}^\Gamma -{\text{\rm Diag}}\, \sigma_{b,\theta},
$$
which is block-wise symmetric.

If $\theta=0$ then $\sigma_{b,0}$ was shown to be separable for each $b>0$ in \cite{kye_osaka}.
On the other hand, if $\theta=\pi$ then $\sigma_{b,\pi}$ was shown \cite{ha+kye_sep} to be separable if and only if $b=1$.
When $b\neq 1$, we note that $\sigma_{b,\pi}$ is nothing but PPT entangled state given by St\o rmer \cite{stormer82} in the early eighties.
We also know that both $\sigma_{b,\theta}$ and $\varrho_{b,\theta}$ are PPT entangled edge states for $0<|\theta|<\frac\pi 3$ by \cite{kye_osaka}.

Now, we turn our attention to the state $\varrho_{b,\pi}$.
We note that $\varrho_{1,\pi}$ is the separable state given by the following four product vectors
$$
\begin{aligned}
&(1,1,1)^{\text{\rm t}} \otimes (1,1,1)^{\text{\rm t}},\, &
&(1,1,-1)^{\text{\rm t}} \otimes (1,1,-1)^{\text{\rm t}},\\
&(1,-1,1)^{\text{\rm t}} \otimes (1,-1,1)^{\text{\rm t}},\,  &
&(-1,1,1)^{\text{\rm t}} \otimes (-1,1,1)^{\text{\rm t}},\\
\end{aligned}
$$
as it was shown in \cite{ha+kye_unique_decom}. The states $\varrho_{b,\pi}$ with $b\neq 1$ appear in the construction \cite{ha+kye}
of PPT entangled states of type $(4,4)$ using the duality between positive linear maps and separable states.
The special case $\varrho_{2,\pi}$ is just the first example
of $3\ot 3$ PPT entangled state given by Choi \cite{choi-ppt}. In short, we see that $\varrho_{b,\pi}$ is separable if and only if $b=1$.

If we take the  diagonal unitary $U={\text{\rm Diag}}(1,e^{-\frac 23\pi i},e^{\frac 23\pi i})$, then we have
$$
U^{-1}
\left(
\begin{matrix}
p_\theta & -e^{i\theta} & -e^{-i\theta}\\
-e^{-i\theta} & p_\theta & -e^{i\theta}\\
-e^{i\theta} & -e^{-i\theta} & p_\theta
\end{matrix}
\right)U
=
\left(
\begin{matrix}
p_\theta & -e^{i(\theta-\frac23\pi)} & -e^{-i(\theta-\frac23\pi)}\\
-e^{-i(\theta-\frac23\pi)} & p_\theta & -e^{i(\theta-\frac23\pi)}\\
-e^{i(\theta-\frac23\pi)} & -e^{-i(\theta-\frac23\pi)} & p_\theta
\end{matrix}\right),
$$
and so it follows that
\begin{equation}\label{unitary}
(I\ot U)^{-1}\varrho_{b,\theta}(I\ot U)=\varrho_{b,\theta-\frac 23\pi},\qquad
(I\ot U)^{-1}\sigma_{b,\theta}(I\ot U)=\sigma_{b,\theta-\frac 23\pi}.
\end{equation}
Therefore, the separability and PPT properties of $\varrho_{b,\theta}$ and $\sigma_{b,\theta}$
are invariant under the translation of $\theta$ by $\frac 23\pi$,
as well as the types of the states.
Therefore, we have the following:

\begin{theorem}
For states $\sigma_\theta$ and $\varrho_\theta$, we have the following:
\begin{enumerate}
\item[(i)]
If $\theta\neq \frac n3\pi$ for any integer $n$, then $\sigma_{b,\theta}$ and $\varrho_{b,\theta}$ are
PPT entangled edge states of type $(8,6)$ and $(5,5)$, respectively.
\item[(ii)]
If $\theta=\frac n3\pi$ for an even integer $n$,
then $\sigma_{b,\theta}$ and $\varrho_{b,\theta}$ are separable states of type $(8,6)$ and $(5,5)$, respectively.
\item[(iii)]
If $\theta=\frac n3\pi$ for an odd integer $n$ and $b=1$,
then $\sigma_{b,\theta}$ and $\varrho_{b,\theta}$ are separable states of type $(7,6)$ and $(4,4)$, respectively.
\item[(iv)]
If $\theta=\frac n3\pi$ for an odd integer $n$ and $b\neq 1$,
then $\sigma_{b,\theta}$ and $\varrho_{b,\theta}$ are PPT entangled states of type $(7,6)$ and $(4,4)$, respectively.
\end{enumerate}
\end{theorem}

The separability and entangledness of the states $\varrho_{b,\theta}$ and $\sigma_{b,\theta}$ are summarized in
Figure 1.
\begin{figure}[t]
\begin{center}
\includegraphics[scale=0.5]{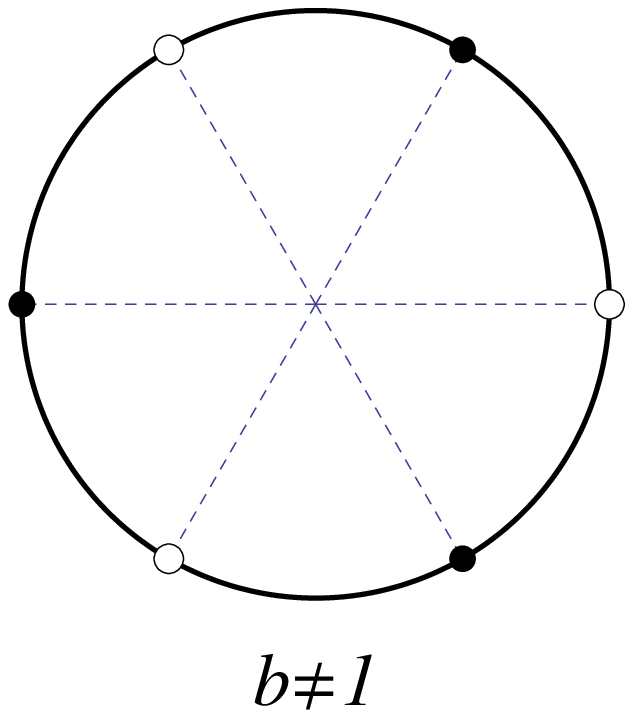}\hskip 2truecm
\includegraphics[scale=0.5]{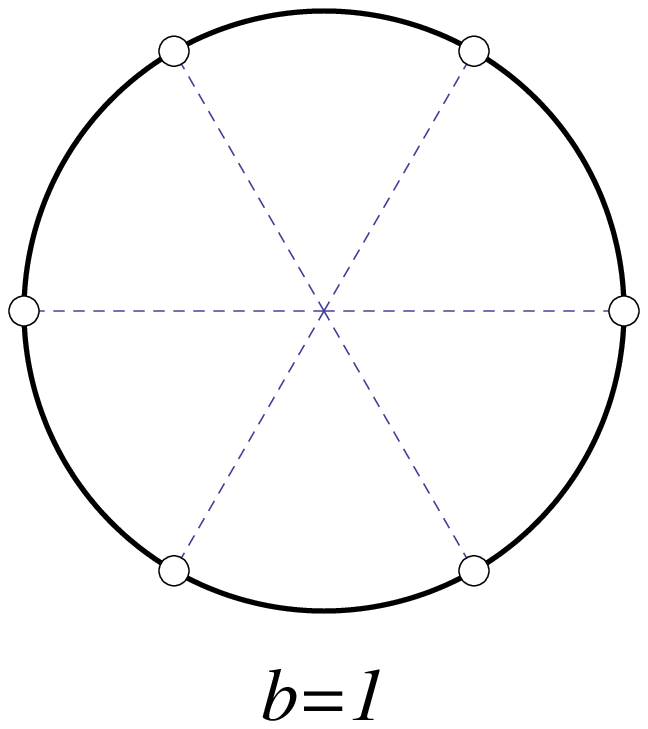}
\end{center}
\caption{The points on the arcs represent PPT entangled states, and the small circles represent separable states. }
\end{figure}
We note that the circle $\{e^{i\theta}:\theta\in\mathbb R\}$ is divided by three arcs by the range of the variable $\theta$:
$$
\left(-\pi,\,-\frac \pi 3\right),\qquad \left(-\frac\pi 3,\,\frac\pi 3\right),\qquad \left(\frac\pi 3,\,\pi\right).
$$
We also note that the following three vectors
$$
\begin{aligned}
w_1(\theta)=&(0,b,0\,;\,e^{i\theta},0,0\,;\,0,0,0),\\
w_2(\theta)=&(0,0,0\,;\,0,0,b\,;\,0,e^{i\theta},0),\\
w_3(\theta)=&(0,0,e^{i\theta}\,;\,0,0,0\,;\,b,0,0),
\end{aligned}
$$
belong to the kernel of $\varrho_{b,\theta}$, regardless of the values of $b$ and $\theta$. There are extra kernel vectors:
$$
\begin{aligned}
w_-&=(1,0,0\,;\,0,e^{\frac 23\pi i},0\,;\,0,0,e^{-\frac 23\pi i}),\qquad &-\pi<&\theta<-\frac\pi 3,\\
w_0&=(1,0,0\,;\,0,1,0\,;\,0,0,1),\qquad &-\frac\pi 3<&\theta<+\frac\pi 3,\\
w_+&=(1,0,0\,;\,0,e^{-\frac 23\pi i},0\,;\,0,0,e^{\frac 23\pi i}),\qquad &\frac\pi 3<&\theta<\pi.
\end{aligned}
$$

If we take $(b,\theta)$ and $(c,\tau)$ so that $e^{i\theta}$ and $e^{i\tau}$ belong to the different arcs, then it is clear that
the kernels of $\varrho_{b,\theta}$ and $\varrho_{c,\tau}$ have the trivial intersection. This means that
the nontrivial convex combination $\rho$ of these two states has the full range space, as well as the partial conjugate.
Therefore, we conclude that this PPT state $\rho$ belongs to the interior of the convex set $\mathbb T_{3,3}$.
In the next section, we will show that each state
$\varrho_{b,\theta}$ is an extreme point in the convex set
$\mathbb T_{3,3}$ consisting of all $3\ot 3$ PPT states, whenever $\theta\neq \frac n3\pi$ for an integer $n$.
From this, we conclude that $\nu(\mathbb T_{3,3})=2$.
If we take $(b,\theta)$ and $(c,\tau)$ so that $e^{i\theta}$ and $e^{i\tau}$ belong to the same arc, then
we note that the convex combinations
of $\varrho_{b,\theta}$ and $\varrho_{c,\tau}$ are on the boundary.

For a given $e^{i\theta}$, we take the antipodal point $e^{i(\theta+\pi)}=-e^{i\theta}$ then we see that
$$
\dfrac12(\varrho_{b,\theta}+\varrho_{c,\theta+\pi})
$$
is a diagonal matrix, and so it is separable. Actually, it is an interior point of $\mathbb S_{3,3}$,
since there is no zero entry in the diagonal.
This shows that the convex combination of two extreme PPT states may be in the interior of the convex set $\mathbb S_{3,3}$
of all separable states. We note that $p_\theta+p_{\theta+\pi}>2$ for each $\theta$,
and so we may take $b>0$ so that $b+\frac 1b=p_\theta+p_{\theta+\pi}$.
Then we see that the sum $\varrho_{b,\theta}+\varrho_{1/b,\theta+\pi}$ of two extreme PPT states is a scalar multiple of the identity matrix.

\section{Extremeness}
First, we briefly explain the method \cite{leinass_mo, ha_ext,augusiak_gkl} to check if a given face $\tau(D,E)$
is an extreme point or not,
where $D$ and $E$ are subspaces of $\mathbb C^m\otimes \mathbb C^n$. Let $(M_m\otimes M_n)_h$ be the real Hilbert space of all
$mn\times mn$ hermitian matrices in $M_m\otimes M_n$ with the inner product $\langle X,Y\rangle=\text{Tr}(YX^{\rm t})$,
and orthogonal projections $P_D$ and $P_E$ in
$(M_m\otimes M_n)_h$ onto $D$ and $E$, respectively. We define real linear maps $\phi_D$ and $\phi_E$ between $(M_m\otimes M_n)_h$ by
\[
\phi_D(X)=P_D X P_D-X,\quad \phi_E(X)=(P_E X^{\Gamma}P_E)^{\Gamma}-X, \quad X\in (M_m\otimes M_n)_h.
\]
Then we see that $\tau(D,E)\subset \text{Ker} \,\phi_D \cap \text{Ker}\,\phi_E$, where $\text{Ker}\,\phi_D$
denotes the kernel space of $\phi_D$. Therefore,
if $\text{Ker} \,\phi_D \cap \text{Ker}\,\phi_E$ is one-dimensional then $\tau(D,E)$ must be an extreme
point. It is not so difficult to see that the converse does hold.
Thus, we can conclude that $\tau(D,E)$ is an extreme point if and only if the condition
\[
\text{dim}(\text{Ker} \,\phi_D \cap \text{Ker}\,\phi_E)=1
\]
holds.

Now, we proceed to show that $\varrho_{b,\theta}$ is an extreme point in the convex set $\mathbb T_{3,3}$,
whenever $0<|\theta|<\frac \pi 3$.  Let $D=\mathcal R \varrho_{b,\theta}$ and $E=\mathcal R \varrho_{b,\theta}^{\Gamma}$.
We  note that $\varrho_{b,\theta}=\varrho_{b,\theta}^{\Gamma}$, so we see that $P_D=P_E$.
Applying the Gram-Schmidt process to linearly independent vectors of $\mathcal R \varrho_{b,\theta}$,
we can compute the orthogonal projection $P_D$ as follows:
\[
P_D=P_E=\begin{pmatrix}
\frac 23 & 0 & 0 & 0 & -\frac 13 & 0 & 0 & 0 & -\frac 13\\
0 &\frac 1{1+b^2} & 0 &-\frac{b e^{-i \theta}}{1+b^2} & 0 & 0 & 0 & 0 & 0\\
0 & 0 & \frac{b^2}{1+b^2} & 0 & 0 & 0 &-\frac{b e^{i \theta}}{1+b^2} & 0 & 0\\
0 & -\frac{b e^{i \theta}}{1+b^2} & 0 & \frac{b^2}{1+b^2} & 0 & 0 & 0 & 0 & 0\\
-\frac 13 & 0 & 0 & 0 & \frac 23 & 0 & 0 & 0 & -\frac 13\\
0 & 0 & 0 & 0 & 0 &\frac 1{1+b^2} & 0 &-\frac{b e^{-i \theta}}{1+b^2} & 0 \\
0 & 0 & -\frac{b e^{-i \theta}}{1+b^2} & 0 & 0 & 0 & \frac{1}{1+b^2} & 0 & 0 \\
0 & 0 & 0 & 0 & 0 &-\frac{b e^{i \theta}}{1+b^2} & 0 & \frac{b^2}{1+b^2} & 0 \\
-\frac 13 & 0 & 0 & 0 & -\frac 13 & 0 & 0 & 0 & \frac 23\\
\end{pmatrix}.
\]
By a direct computation, we can show that both $\text{Ker}\,\phi_D$ and $\text{Ker}\, \phi_E$ are twenty-five dimensional real linear subspaces.
Let $\{E_{ij}\}$ be the usual matrix units in $M_9$. Then,
we can find a basis $\{X_i:1\le i\le 25\}$ of real linear space $\text{Ker}\,\phi_D$,
which consists of hermitian matrices including the following vectors:
\[
\begin{aligned}
X_1&=E_{11}+E_{55}-E_{15}-E_{51}\\
X_2&=E_{11}+E_{99}-E_{19}-E_{91}\\
X_3&=E_{55}+E_{99}-E_{59}-E_{95}\\
X_4&=i (E_{19}-E_{15}-E_{59})-i(E_{91}-E_{51}-E_{95})\\
X_{5}&=e^{-i\theta}E_{24}+e^{i\theta}E_{42}-bE_{44}-\frac 1b E_{22},\\
X_{6}&=e^{-i\theta}E_{68}+e^{i\theta}E_{86}-bE_{88}-\frac 1b E_{66},\\
X_{7}&=e^{-i\theta}E_{73}+e^{i\theta}E_{37}-bE_{33}-\frac 1b E_{77}.
\end{aligned}
\]
We also see that  $\text{Ker}\, \phi_E=\text{span}\{Y_i:1\le i \le 25\}$ with hermitian matrices $Y_i$'s.
Here, we just write down the list of $Y_i$ for $i=1,2,\cdots,7$, as follows:
\[
\begin{aligned}
Y_{1}&=E_{11}+E_{55}-E_{24}-E_{42},\\
Y_{2}&=E_{11}+E_{99}-E_{37}-E_{73},\\
Y_{3}&=E_{55}+E_{99}-E_{68}-E_{86},\\
Y_{4}&=i(E_{37}+E_{42}+E_{86})-i(E_{73}+E_{24}+E_{68}),\\
Y_{5}&=e^{-i\theta}E_{19}+e^{i\theta}E_{91}-bE_{33}-\frac 1 bE_{77},\\
Y_{6}&=e^{-i\theta}E_{51}+e^{i\theta}E_{15}-bE_{44}-\frac 1 bE_{22},\\
Y_{7}&=e^{-i\theta}E_{95}+e^{i\theta}E_{59}-bE_{88}-\frac 1 bE_{66}.
\end{aligned},
\]
For the full list of vectors $X_i$ and $Y_i$ for $8\le i \le 25$, see the appendix.
By solving the linear equation $\sum_{i=1}^{25} x_i X_i=\sum_{j=1}^{25}y_j Y_j$ with respect to $x_i$'s and $y_j$'s, we see that the subspace
$\text{Ker}\, \Phi_D\cap \text{Ker}\, \Phi_E$ is generated $\varrho_{b,\theta}$.
In fact, we have
\[
\begin{aligned}
\varrho_{b,\theta}&=\cos\theta(X_1+X_2+X_3)+\sin\theta X_4-X_5-X_6-X_7\\
&=\cos\theta(Y_1+Y_2+Y_3)-\sin\theta Y_4-Y_5-Y_6-X_7.
\end{aligned}
\]
Therefore, we see that $\varrho_{b,\theta}$ is an extreme point in $\mathbb T_{3,3}$ for $0<|\theta|<\frac {\pi}3$.

We note that $\varrho_{b,\theta-\frac 23 \pi}$ is  extreme  if and only if $\varrho_{b,\theta}$ is so by the relation \eqref{unitary}.
Consequently, we may conclude that $\varrho_{b,\theta}$ is extreme whenever $\theta\neq \frac n3\pi$ for an integer $n$.

\section{Decomposable and positive maps}

In order to see that $\nu(\mathbb P_{3,3})=2$, we recall the
positive linear map $\Phi_{\theta}(t)$ considered in
\cite{ha+kye_exposed}, which maps a $3\times 3$ matrix $X=(x_{ij})$
to the following $3\times 3$ matrix
\[
\begin{pmatrix}
a(t) x_{11}+b(t) x_{22}+c(t) x_{33} & -e^{i\theta} x_{12} & -e^{-i\theta}x_{13}\\
-e^{-i\theta}x_{21} & c(t)x_{11}+a(t)x_{22}+b(t)x_{33} & -e^{i\theta}x_{23}\\
-e^{i\theta}x_{31} &-e^{-i\theta}x_{32} & b(t)x_{11}+c(t) x_{22}+a(t) x_{33}
\end{pmatrix},
\]
where
\[
a(t)=1-\frac{(p_{\theta}-1)t}{1-t+t^2},\quad b(t)=\frac{(p_{\theta}-1)t^2}{1-t+t^2},\quad
c(t)=\frac {(p_{\theta}-1)}{1-t+t^2},
\]
with $0<t<\infty$. It was shown that $\Phi_{\theta}(t)$ generates an exposed ray of the convex cone $\mathbb P_{3,3}$,
and so generates an extreme ray of $\mathbb P_{3,3}$,
whenever the condition
$$
\theta \neq\frac {2n-1}3\pi,\qquad  (\theta, t)\neq \left(\frac {2n}3\pi,\, 1\right)
$$
holds. It is now clear that if we take the convex combination of two antipodal maps $\Phi_\theta(t)$ and $\Phi_{\theta+\pi}(s)$
then we get a positive map whose Choi matrix is a
diagonal matrix with positive diagonal entries, and so we see that this map is an interior point of $\mathbb P_{3,3}$ by duality.

It remains to consider the convex cone $\mathbb D_{m,n}$ consisting of all decomposable maps from $M_m$ into $M_n$. We first note that
every decomposable map is the convex combination of the maps
$$
\phi_V:X\mapsto V^*XV,\qquad \phi^W: X\mapsto W^*X^\ttt W,\qquad X\in M_m,
$$
for $m\times n$ matrices $V$ and $W$, where $X^\ttt$ denotes the transpose of $X$.
Therefore, every decomposable map from $M_m$ into $M_n$ is of the form
\begin{equation}\label{decom}
\phi_{\mathcal V}+\phi^{\mathcal W}= \sum_i \phi_{V_i}+\sum_j \phi^{W_j},
\end{equation}
for a finite sets $\mathcal V=\{V_i\}$ and $\mathcal W=\{W_j\}$ of $m\times n$ matrices.
We also note that
if the map (\ref{decom}) is on the boundary of the cone $\mathbb P_{m,n}$ then it is also on the boundary of
the cone $\mathbb D_{m,n}$. For a product vector $|z\rangle=|\xi\rangle \otimes |\eta\rangle$,
the pairing in \eqref{pairing} is given by
$$
\langle |z\rangle \langle z|,\phi_{\mathcal V}+\phi^{\mathcal W}\rangle
=\sum_i \left| \langle \xi |V_i|\bar \eta\rangle\right|^2+\sum_j |\langle \bar \xi| W_j|\bar \eta\rangle |^2.
$$
Therefore, the map (\ref{decom}) is on the boundary of $\mathbb P_{m,n}$ if and only if the equation
$$
\langle \xi |V_i|\bar \eta\rangle=0,\qquad \langle \bar \xi| W_j|\bar \eta\rangle=0
$$
has a common solution $|\xi\rangle\otimes |\eta\rangle \in\mathbb C^m\ot\mathbb C^n$.
If we put $k=\dim\spa{\mathcal V}$ and $\ell=\dim\spa{\mathcal W}$ then
it was shown in \cite{kye-prod-vec} that
\begin{enumerate}
\item[(i)]
If $k+\ell < m+n-2$, then there exists a
solution
\item[(ii)]
If $k+\ell = m+n-2$ and
$$
\sum_{r+s=m-1}(-1)^r \binom kr\binom \ell s \neq 0,
$$
then there exists a solution.
\item[(iii)]
If $k+\ell > m+n-2$, then the existence of solutions is not guaranteed.
\end{enumerate}
Therefore, we have the following:

\begin{theorem}\label{ukyilhg}
For given natural numbers $m,n=2,3,\dots$, consider the equation
\begin{equation}\label{Krawtchouk}
k+\ell = m+n-2,\qquad \sum_{r+s=m-1}(-1)^r \binom kr\binom \ell s = 0
\end{equation}
with unknowns $k$ and $\ell$. Then, we have the following:
\begin{enumerate}
\item[(i)]
We have $\nu(\mathbb D_{m,n})\ge m+n-2$ in general.
\item[(ii)]
If the equation {\rm (\ref{Krawtchouk})} has no solution then $\nu(\mathbb D_{m,n})\ge m+n-1$.
\end{enumerate}
\end{theorem}

The polynomial in the Diophantine equation (\ref{Krawtchouk}) is called the Krawtchouk polynomial
which plays an important role in coding theory. See \cite{MWS} and \cite{vint}. The
equation (\ref{Krawtchouk}) has not yet completely solved.

In order to get an upper bound for $\nu(\mathbb D_{m,n})$, we have to construct decomposable maps
in the interior of the cone $\mathbb D_{m,n}$. By the duality
between decomposable maps and PPT states with respect to the pairing \eqref{pairing},
we see that the map in (\ref{decom}) lies on the boundary of the cone $\mathbb D_{m,n}$
if and only if there exists a PPT states $\sigma$ such that the ranges
of $\sigma$ and the partial transpose $\sigma^\Gamma$ coincide with
${\mathcal V}^\perp$ and ${\mathcal W}^\perp$, respectively.

We consider the case with $m=2$, to take an $n-1$ dimensional subspace $D$ of $\mathbb C^2\ot \mathbb C^n$
with no product vectors \cite{wallach,parth}. Then it is clear that there is no PPT state $\sigma$ such that ${\mathcal R}\sigma=D$ and
${\mathcal R}\sigma^\Gamma=\mathbb C^2\ot\mathbb C^n$. Indeed, if we assume that there is such a state $\sigma$ then $\sigma$ must be separable
by \cite{2xn}, but this state violates the range criterion \cite{p-horo}. Therefore, if we take a basis ${\mathcal V}$
in $D^\perp$ then the map $\phi_{\mathcal V}$ is an interior point of the cone $\mathbb D_{2,n}$. This shows that
$\nu(\mathbb D_{2,n})\le n+1$. In the case of $m=2$, it was shown in \cite{kye-prod-vec} that the equation (\ref{Krawtchouk}) has a solution
if and only if $n$ is an even number. This proves the odd case of the following:
\begin{equation}\label{2xxn}
\nu(\mathbb D_{2,n})=
\begin{cases}
n+1,\quad &n\ {\text{\rm is odd}},\\
n,\quad &n\ {\text{\rm is even}}.
\end{cases}
\end{equation}
When $n=2\mu$ is an even integer then the equation (\ref{Krawtchouk}) has the unique
solution $(k,\ell)=(\mu,\mu)$, and one can construct
${\mathcal V}=\{V_1,\dots,V_{\mu}\}$ and ${\mathcal W}=\{W_1,\dots,W_{\mu}\}$ so that the decomposable map
(\ref{decom}) is an interior point of $\mathbb D_{2,n}$,
following the argument in \cite{kye-prod-vec}.
To do this, we consider the $2\times 2\mu$ matrix $V_i$ whose $i$-th $2\times 2$ block is the identity
matrix and other entries are all zero. We also consider the $2\times 2\mu$ matrix
$W_i$ whose $i$-th block is $\left(\begin{matrix}0&-1\\1&0\end{matrix}\right)$ and other entries are all zero. Then we see that
$$
\sum_{i=1}^\mu\phi_{V_i}+\sum_{i=1}^\mu\phi^{W_i}
$$
is just the trace map sending $X\in M_2$ to $\text{\rm Tr}(X) I \in M_{2\mu}$,
which is an interior point of $\mathbb D_{2,2\mu}$. This shows the above equality (\ref{2xxn}) when $n$ is even.
For the $2\otimes 2$ system, the whole facial structures of the cone $\mathbb D_{2,2}$
have been characterized in \cite{byeon-kye}.

In the case of $m=3$, we know that the equation (\ref{Krawtchouk}) has a solution if and only if $n$ is of the form $n=\mu(\mu+2)$,
with the solution $(k,\ell)=(\binom{\mu+1} 2,\binom{\mu+2}2)$. Especially, in the $3\ot 3$ case, we have the solution $(k,\ell)=(1,3)$.
In this case, we see that the map
$$
\phi_I+\phi^{E_{12}-E_{21}}+\phi^{E_{23}-E_{32}}+\phi^{E_{31}-E_{13}}
$$
is exactly the trace map, which is an interior point of $\mathbb D_{3,3}$. Therefore, we have
$$
\nu(\mathbb D_{3,3})=4.
$$

\section{Discussion}
For a given convex set, we have considered the smallest number of extreme points with which we may get
an interior point by their convex combinations. For $3\otimes 3$ PPT states and positive maps, these numbers turned out to be just $2$.
This means that there exist \lq antipodal\rq\ extreme points. Poor knowledge on extreme PPT states and extreme positive maps prevent the authors
to extend these results to higher dimensions.

For the cases of separable states and decomposable maps, these numbers exceed $2$. This means that there exist no
\lq antipodal\rq\ extreme points, and might reflect the facts that the notions of separability and decomposability are defined by
convex hulls of prescribed given extreme points, and that there are no easy intrinsic characterizations for these notions.

The equality $\nu(\mathbb S_{m,n})=mn$ tells us that the number $mn$
is the minimum of the lengths of interior points of $\mathbb
S_{m,n}$. Recall that the length of a separable state $\varrho$ is
given by the minimum number of product states with which $\varrho$
can be expressed as a convex combination. It seems to be an
interesting question to ask if every interior point of $\mathbb
S_{m,n}$ has the length $mn$. The authors
\cite{ha+kye_unique_decom,ha+kye_2x4} have recently constructed
separable states in $\mathbb S_{m,n}$ whose lengths exceed the
number $mn$, for the cases $(m,n)=(3,3)$ and $(2,4)$. All of those
are boundary points of the convex set $\mathbb S_{m,n}$. See also \cite{chen_dj_semialg}. For some
faces of $\mathbb S_{m,n}$, it is possible to characterize the
interior by lengths. For example, this is clearly the case if a face
of $\mathbb S_{m,n}$ is affinely isomorphic to a simplex. See
\cite{ha+kye_unique_decom,ha+kye_2x4} for constructions of such
faces in the $3\otimes 3$ or $2\otimes n$ cases. This is also the
case \cite{kye_trigono} for a face of $\mathbb S_{2,n}$ which is
affinely isomorphic to the convex set generated by trigonometric
moment curve.


\section{Appendix}
In this appendix we list up
the remaining vectors $X_i$ and $Y_i$ constituting bases of real spaces
$\text{Ker}(\phi_D)$ and $\text{Ker}(\phi_E)$, respectively, as follows:

\[
\begin{aligned}
X_{8}&=e^{-i\theta}(E_{29}-E_{21})+e^{i\theta}(E_{92}-E_{12})+b(E_{14}+E_{41}-E_{49}-E_{94}),\\
X_{9}&=e^{-i\theta}(E_{71}-E_{75})+e^{i\theta}(E_{17}-E_{57})+b(E_{35}+E_{53}-E_{13}-E_{31}),\\
X_{10}&=e^{-i\theta}(E_{79}-E_{71})+e^{i\theta}(E_{97}-E_{17})+b(E_{13}+E_{31}-E_{39}-E_{93}),\\
X_{11}&=e^{-i\theta}(E_{61}-E_{65})+e^{i\theta}(E_{16}-E_{56})+b(E_{58}+E_{85}-E_{18}-E_{81}),\\
X_{12}&=e^{-i\theta}(E_{69}-E_{61})+e^{i\theta}(E_{96}-E_{16})+b(E_{18}+E_{81}-E_{89}-E_{98}),\\
X_{13}&=-e^{-i\theta}(E_{21}+E_{25})-e^{i\theta}(E_{12}+E_{52})+b(E_{14}+E_{41}-E_{45}-E_{54}),\\
X_{14}&=e^{-i\theta}(E_{13}-E_{53})+e^{i\theta}(E_{31}-E_{35})+\frac 1 b(E_{57}+E_{75}-E_{17}-E_{71}),\\
X_{15}&=e^{-i\theta}(E_{13}-E_{93})+e^{i\theta}(E_{31}-E_{39})+\frac 1 b(E_{97}+E_{79}-E_{17}-E_{71}),\\
X_{16}&=e^{-i\theta}(E_{14}-E_{54})+e^{i\theta}(E_{41}-E_{45})+\frac 1 b(E_{25}+E_{52}-E_{12}-E_{21} ),\\
X_{17}&=e^{-i\theta}(E_{14}-E_{94})+e^{i\theta}(E_{41}-E_{49})+\frac 1 b(E_{29}+E_{92}-E_{12}-E_{21} ),\\
X_{18}&=e^{-i\theta}(E_{18}-E_{98})+e^{i\theta}(E_{81}-E_{89})+\frac 1 b(E_{69}+E_{96}-E_{16}-E_{61} ),\\
X_{19}&=e^{-i\theta}(E_{58}-E_{18})+e^{i\theta}(E_{85}-E_{81})+\frac 1 b(E_{16}+E_{61}-E_{56}-E_{65} ),\\
X_{20}&=e^{-i\theta}(E_{63}+E_{78})+e^{i\theta}(E_{36}+E_{87})-b(E_{38}+E_{83})-\frac 1 b(E_{67}+E_{76} ),\\
X_{21}&=-e^{-i\theta}(E_{23}+E_{74})-e^{i\theta}(E_{32}+E_{47})+b(E_{34}+E_{43})+\frac 1 b(E_{27}+E_{72} ),\\
X_{22}&=-e^{-i\theta}(E_{28}+E_{64})-e^{i\theta}(E_{82}+E_{46})+b(E_{48}+E_{84})+\frac 1 b(E_{26}+E_{62} ),\\
X_{23}&=e^{-i\theta}(E_{67}+b^2E_{83}-be^{-i\theta}E_{63})+e^{i\theta}(E_{76}+b^2E_{38}-be^{i\theta}E_{36})-b(E_{78}+E_{87}),\\
X_{24}&=e^{-i\theta}(E_{48}+\frac 1{b^2}E_{26}-\frac 1 b e^{-i\theta}E_{28})+e^{i\theta}(E_{84}+\frac 1{b^2}E_{62}-\frac 1{b}e^{i\theta}E_{82})-\frac 1b(E_{46}+E_{64}),\\
X_{25}&=-e^{-i\theta}(E_{43}+\frac 1{b^2}E_{27}-\frac 1 b e^{-i\theta}E_{23})-e^{i\theta}(E_{34}+\frac 1{b^2}E_{72}-\frac 1{b}e^{i\theta}E_{32})+\frac 1b(E_{47}+E_{74}),\\
\end{aligned}
\]
\[
\begin{aligned}
Y_{8}&=e^{-i\theta}(E_{21}-E_{83})+e^{i\theta}(E_{12}-E_{38})+b(E_{67}+E_{76}-E_{14}-E_{41}),\\
Y_{9}&=e^{-i\theta}(E_{21}-E_{52})+e^{i\theta}(E_{12}-E_{25})+b(E_{45}+E_{54}-E_{14}-E_{41}),\\
Y_{10}&=e^{-i\theta}(E_{34}-E_{65})+e^{i\theta}(E_{43}-E_{56})+b(E_{58}+E_{85}-E_{27}-E_{72}),\\
Y_{11}&=e^{-i\theta}(E_{34}-E_{96})+e^{i\theta}(E_{43}-E_{69})+b(E_{89}+E_{98}-E_{27}-E_{72}),\\
Y_{12}&=e^{-i\theta}(E_{48}-E_{17})+e^{i\theta}(E_{84}-E_{71})+b(E_{13}+E_{31}-E_{26}-E_{62}),\\
Y_{13}&=e^{-i\theta}(E_{79}-E_{17})+e^{i\theta}(E_{97}-E_{71})+b(E_{13}+E_{31}-E_{39}-E_{93}),\\
Y_{14}&=e^{-i\theta}(E_{26}-E_{13})+e^{i\theta}(E_{62}-E_{31})+\frac 1b(E_{17}+E_{71}-E_{48}-E_{84}),\\
Y_{15}&=e^{-i\theta}(E_{39}-E_{13})+e^{i\theta}(E_{93}-E_{31})+\frac 1b(E_{17}+E_{71}-E_{79}-E_{97}),\\
Y_{14}&=e^{-i\theta}(E_{26}-E_{13})+e^{i\theta}(E_{62}-E_{31})+\frac 1b(E_{17}+E_{71}-E_{48}-E_{84}),\\
Y_{15}&=e^{-i\theta}(E_{39}-E_{13})+e^{i\theta}(E_{93}-E_{31})+\frac 1b(E_{17}+E_{71}-E_{79}-E_{97}),\\
Y_{14}&=e^{-i\theta}(E_{26}-E_{13})+e^{i\theta}(E_{62}-E_{31})+\frac 1b(E_{17}+E_{71}-E_{48}-E_{84}),\\
Y_{15}&=e^{-i\theta}(E_{39}-E_{13})+e^{i\theta}(E_{93}-E_{31})+\frac 1b(E_{17}+E_{71}-E_{79}-E_{97}),\\
Y_{16}&=e^{-i\theta}(E_{41}-E_{54})+e^{i\theta}(E_{14}-E_{45})+\frac 1b(E_{25}+E_{52}-E_{12}-E_{21}),\\
Y_{17}&=e^{-i\theta}(E_{67}-E_{41})+e^{i\theta}(E_{76}-E_{14})+\frac 1b(E_{12}+E_{21}-E_{38}-E_{83}),\\
Y_{18}&=e^{-i\theta}(E_{72}-E_{85})+e^{i\theta}(E_{27}-E_{58})+\frac 1b(E_{56}+E_{65}-E_{34}-E_{43}),\\
Y_{19}&=e^{-i\theta}(E_{72}-E_{98})+e^{i\theta}(E_{27}-E_{89})+\frac 1b(E_{69}+E_{96}-E_{34}-E_{43}),\\
Y_{20}&=e^{-i\theta}(E_{36}+E_{78})+e^{i\theta}(E_{63}+E_{87})-b(E_{29}+E_{92})-\frac 1b (E_{49}+E_{94}),\\
Y_{21}&=-e^{-i\theta}(E_{64}+E_{82})-e^{i\theta}(E_{46}+E_{28})+b(E_{57}+E_{75})+\frac 1b (E_{35}+E_{53}),\\
Y_{22}&=e^{-i\theta}(be^{-i\theta}E_{36}-E_{94}-b^2 E_{29})+e^{i\theta}(b e^{i\theta}E_{63}-E_{49}-b^2 E_{92})+b(E_{78}+E_{87}),\\
Y_{23}&=e^{-i\theta}(\frac {e^{-i\theta}} b E_{82}-\frac 1{b^2}E_{53}-E_{75})+e^{i\theta}(\frac {e^{i\theta}}b E_{28}-\frac 1{b^2}E_{35}-E_{57})+\frac 1 b(E_{46}+E_{64}),\\
Y_{24}&=e^{-i\theta}(e^{-i\theta} E_{23}-bE_{16}-\frac 1b E_{81})+e^{i\theta} (e^{i\theta} E_{32}-b E_{61}-\frac 1b E_{18})+(E_{47}+E_{74}),\\
Y_{25}&=e^{-i\theta}(e^{-i\theta} E_{47}-bE_{61}-\frac 1b E_{18})+e^{i\theta} (e^{i\theta} E_{74}-b E_{16}-\frac 1b E_{81})+(E_{23}+E_{32}).\\
\end{aligned}
\]

\end{document}